\documentstyle[12pt,here,rotate,epsf]{article}
\newcommand{\bel}[1]{\begin{equation}\label{#1}}
\newcommand{\ee}{\end{equation}}

\author{H.~Hofmann 
\thanks{e-mail: hhofmann@.physik.tu-muenchen.de}
\thanks{http://www.physik.tu-muenchen.de/tumphy/e/T36/hofmann.html}\\
        Physik-Department, TU M\"unchen, D-85747 Garching\\
        and\\
        D.~Kiderlen\\
        National Superconducting Cyclotron Laboratory\\
        Michigan State University, East Lansing, Michigan 48824\\
        Physik-Department, TU M\"unchen, D-85747 Garching
        \vspace{0.8cm}}

\title{Statistical fluctuations for the fission process on its decent from
saddle to scission}

\begin{document}
\maketitle

\begin{abstract}
\noindent
We reconsider the importance of statistical fluctuations for
fission dynamics beyond the saddle in the light of recent evaluations of
transport coefficients for average motion. The size of these
fluctuations are estimated by means of the Kramers-Ingold solution for
the inverted oscillator, which allows for an inclusion of quantum effects.
\end{abstract}

\noindent
 \vspace*{1cm}
\centerline{PACS: 5.40+j, 5.60+w, 24.10 Pa, 24.60.-k}
 \vspace*{1cm}

Kramers' transport equation \cite{kram} delivers the classic description
of the dynamics of fission at finite excitation. It restricts to the
high temperature regime in which collective motion is treated within
the frame of classical statistical mechanics. An extension to the
quantum case has been described in \cite{hosaoc}, \cite{hofing} and
\cite{hofthoming}. This is possible within a locally harmonic
approximation, in which global motion is described in terms of local
propagators. The latter are constructed as special solutions of an
appropriate transport equation in which anharmonic forces are
linearized around some given point in phase space \cite{hofrep}. For the
present purpose this equation is needed only for the region between
barrier and scission point. To simplify matters we like to apply
the schematic model of \cite{hofnix}, in which the potential is
represented by an inverted oscillator (centered at $q=0$) 
and where the coefficients for inertia $M$ and friction $\gamma$ do not
change along the fission path. The transport equation for the
distribution $ d(q,P,t)$ in collective phase space may then be written
as:
\bel{traneq}
{\partial  \over \partial t} d(q,P,t) = \Biggl[
    -{\partial\over\partial q}{P\over M} + {\partial\over\partial P}Cq
    +{\partial\over\partial P}\gamma {P\over M}
    + \nonumber \\
     D_{qp}{\partial^2\over\partial q\partial P} + D_{pp}
    {\partial^2\over\partial P\partial P} \Biggr] d(q,P,t)
\ee
with $C$ being the (negative) stiffness coefficient. There is little
doubt that a correct treatment would require to go beyond the harmonic
approximation underlying the form (\ref{traneq}). Not only that the 
potential energy along the fission path will, in general, be more 
complicated than given by the simple quadratic dependence $C q^2/2$.
Also the other transport coefficients will vary with $q$. However,
recent numerical computations \cite{yaivho}, \cite{ivhopaya} have
revealed that the following ratios 
\bel{effcoef}
\varpi= \sqrt{\mid C \mid \over M} \qquad\qquad
\eta= {\gamma \over 2 M \varpi}
\ee
are quite stable. As we shall see soon, within the harmonic
approximation it is them which parameterize the quantities we are
mostly interested in, namely the kinetic energy and its fluctuation.

As compared to the form of Kramers' equation, only two modifications
occur in (\ref{traneq}), both referring to the diffusive terms.
Firstly, there appears a cross term with the $D_{qp}$ being different
from zero in the quantal regime. Secondly, the coefficient $D_{pp}$
will be given by the classic Einstein relation  $D_{pp}= \gamma T$ only at
large temperatures when quantum effects disappear. In their quantum
version, these coefficients are defined by the following expressions
\bel{diffpp}
D_{pp}  = \gamma M\int_{\cal C} \! {{\rm d}\omega \over 2\pi }\,
  \hbar \coth \left( {\hbar\omega \over 2T}\right) \, 
  \chi^{\prime\prime}_{qq} (\omega )\,\omega^2 
  \equiv {\gamma \over M} \Sigma^{\rm eq}_{pp}
\ee
\bel{diffqp}
D_{qp}  = \int_{\cal C} \! {{\rm d}\omega \over 2\pi }\,
 \hbar \coth \left( {\hbar\omega \over 2T}\right) \,  
  \chi^{\prime\prime} _{qq} (\omega )
\left[C- M \omega^2 \right]
  \equiv C\Sigma^{\rm eq}_{qq} - {1\over M} \Sigma^{\rm eq}_{pp}
\ee
They are a consequence of the fluctuation dissipation theorem (FDT):
Together with some simple symmetry relations, it allows one to 
calculate the "equilibrium fluctuations" $\Sigma^{\rm eq}_{qq}$ and
$\Sigma^{\rm eq}_{pp}$ from the dissipative part $\chi^{\prime\prime}
_{qq} (\omega )$ of the response function which represents average
motion in $q(t)$. For the linearized version used here, the latter must
be related to the one of a damped oscillator determined by $M,\gamma$
and $C$. If (\ref{traneq}) is applied to a bound oscillator with
$C>0$, the expressions (\ref{diffpp}, \ref{diffqp}) warrant that for
$t\to \infty$ the dynamical fluctuations in $q,P$ turn into those of
equilibrium as determined by the FDT \cite{hosaoc}.  Thus the contour
$\cal C$ has to be chosen in the common way, namely along the real axis
extending from $-\infty$ to $\infty$. For unbound motion with $C<0$,
the case we want to study here, one has to apply suitable analytical
continuations. This is possible in two ways: (i) One may evaluate the
forms given in (\ref{diffpp}, \ref{diffqp}) for $C>0$ and perform the
continuations in the expressions one obtains after performing the
integration (see \cite{hofing} and \cite{hofthoming}). (ii) In an
alternative method \cite{kiderlendoc} one redefines
the contour $\cal C$. This allows for more general applications like in
\cite{kidhofpl}. For more details we refer to \cite{khqstsu} and
\cite{hofrep}.

With respect to the evaluation of (\ref{diffpp}, \ref{diffqp}) we need
to clarify a problem hidden in the integrals. In the quantum regime,
some of them would diverge if we one were to take for the response
function that of the damped oscillator. This problem is well known, and
one possible solution is to apply the Drude regularization.  This means
to replace the $\chi_{qq} (\omega )$ by
\bel{druderes}
\chi_{D}(\omega)= {1\over -M\omega^2 - i \gamma(\omega) \omega + C}
\ee
with a frequency dependent friction coefficient 
\bel{drufri}
{\gamma_D(\omega)\over \gamma } =
   \left(1 - i{\omega \over \varpi_D}\right)^{-1} 
\ee
In this way a "cut-off" frequency $\omega_D$ is introduced. 
We do not want to discuss the interesting questions of how its value
can be fixed and from which physical quantities. For the computations
to be discussed below we chose $\omega _D= 10 \varpi$,  with the
$\varpi$ given by (\ref{effcoef}). Fortunately, the diffusion coefficients do
not depend on $\omega _D$ too much (see \cite{khqstsu}). Changing the
latter by a factor of two, our final results would have to be modified
by less than $30 \%$ which, as we shall see, will not influence much
the conclusions we are going to draw below. 

Notice, please, that this regularization problem disappears in the
classical limit. The latter is obtained if the $ \hbar \coth
\left(\hbar\omega / (2T)\right)$ is replaced by $2T/\omega$. With such
a weighting factor all integrals in (\ref{diffpp}) and (\ref{diffqp})
converge even for the case of a constant friction force. Apparently
this classical limit is identical to the high temperature limit, for
which one needs to have $\hbar\omega\ll T$. Looking back to the right
hand sides of (\ref{diffpp}) and (\ref{diffqp}), it becomes evident why
in this case the diffusion coefficients turn into those of Kramers'
equation. For stable modes this is simply a consequence of the
(classical) equipartition theorem. Since the diffusion coefficients
then become independent of $C$ the analytic continuation is trivial,
and does not change the results for $D_{pp}$ and $D_{qp}$ when turning
to the unstable situation.

Whereas for stable modes the extension of Kramers' equation to the
quantum regime is possible for all temperatures and all possible values
of the transport coefficients, for unstable ones the description
ceases to make sense at very low temperatures. First of 
all, below a certain $T_0$ it is not possible anymore to save the
integral representation of the FDT. The contour $\cal C$ needs to cross
the imaginary axis between the pole $\omega_+=i\mid \omega_+ \mid$ of
the unstable mode and the first Matsubara frequency
$\omega^1_M=2\pi T/\hbar$ (see \cite{khqstsu}-\cite{kidhofpl}). The $T_0$
obtained in this way, namely $T_0=\hbar \omega_+ \mid /(2\pi)$, is
identical to the so called "cross over" temperature known from
treatments of "dissipative tunneling" with functional integrals in the
imaginary time domain. Here we are looking at real time propagation for
which in a model case Ingold \cite{ingold} has been able to construct a
phase space distribution corresponding to a constant flux across the
barrier. Such a construction is possible above a critical temperature
$T_c>T_0$.  As shown in \cite{hofing} this distribution solves the
transport equation (\ref{traneq}) if only the diffusion coefficients
are defined as described above. In this sense it may be considered 
the generalization to the quantum regime of the stationary
solution found by Kramers. Actually, this $T_c$ turns out to be that
temperature at which the $C \Sigma^{\rm eq}_{qq}$ becomes negative. It
is of the order of $0.5~{\rm MeV}$ or less, depends on the transport
coefficients of average motion and decreases with increasing damping, see
\cite{khqstsu}. On the nuclear scale such values of $T$ can be 
considered small. (As a matter of fact, in such a regime the very
concept of temperature itself becomes questionable.) Commonly, the
fission experiments, which one may want to interpret with such a transport
equation, involve higher excitations, for which our extension 
thus applies. Unfortunately, an experimental verification of the
existence of quantum effects is still missing. Two possibilities have
been suggested so far, the (dwell) time  $\tau$ from saddle to scission
\cite{hofing} (see below), and the decay rate \cite{hofthoming}. For those,
however, quantum effects would show up only in a very narrow range
between $T_c$ and values of about $T \simeq 1\cdots 1.5
~{\rm MeV}$.

This situation may change if one looks at quantities which involve the
momentum distribution.  Such a feature is known from studies of the
dynamics of {\it stable modes}. There the equilibrium fluctuations in
the coordinate get squeezed when friction increases. In this way their
values get {\it closer} to the {\it classical} limit. The opposite
holds true for the momentum.  A nice demonstration of this effect can
be found in \cite{graschring} where path integrals are applied to a
solvable model. In \cite{khqstsu} this problem has been taken up for the
nuclear context within the locally harmonic approximation, for which
one is not restricted to describe the "heat bath" of the nucleonic
degrees of freedom by a set of coupled oscillators. 

In the present letter we specifically address the dynamics across the
fission barrier. Different to \cite{hofing} and \cite{hofthoming} we
want to exploit the Kramers-Ingold solution $ d_I(q,P)$ to evaluate the
kinetic energy and its variance at scission. The calculation can be
done in complete analogy to the case discussed in \cite{hofnix} for
Kramers' equation.  The average kinetic energy at scission may be
defined by
\bel{averen}
E_{\rm kin} = \int_{-\infty}^\infty\; \frac{P\,dP}{M j}\; d_I(q_{sc},P)\,
\frac{P^2}{2M}
\ee 
and its variance by
\bel{fluc}
\sigma^2_{\rm kin} = \int_{-\infty}^\infty\; \frac{P\,dP}{M j}\;
d_I(q_{sc},P)\, 
\left(\frac{P^2}{2M}\right)^2 \quad - \quad 
\left(E_{\rm kin} \right)^2.
\ee
with the $q_{sc}$ being the coordinate of the scission point. 
Notice please, that the sampling is done for the following
normalization of the distribution 
\bel{flux}
 \int_{-\infty}^\infty\; \frac{P\,dP}{M j}\; d_I(q,P) =1 
\ee
with $j$ being the constant flux across the barrier.
Like in the classical case (see eqs.(13) and (14) of \cite{hofnix})
the integrals can be carried out analytically. Introducing the
abbreviation 
\bel{zplusnor}
\zeta = \sqrt{1+\eta^2}-\eta \equiv {1\over \sqrt{1+\eta^2}+\eta}
\ee
the result writes as
\bel{averendif}
 E_{\rm kin}= 
    {1\over2}\left( M\varpi^2q_{sc}^2 - D_{qp}\right)
  \zeta^2 
     + {D_{pp}\over\gamma}\left(1 + \eta\zeta\right) 
\ee
and
\bel{enerfluctdif}  \sigma^2=  
    \left( { D_{pp}\over\gamma } \right)^2 +
         {1\over 2}\zeta^2 \left[ 
              \left( 2\eta {D_{pp}\over\gamma}- \zeta D_{qp} +
              M\varpi^2 q_{sc}^2 \zeta\right)^2 -
             ( M\varpi^2 q_{sc}^2 \zeta )^2  \right]
\ee
For the difference $\Delta V={1\over2} M\varpi^2q_{sc}^2$ of the potential
energy between saddle and scission the value $\Delta V=20~{\rm MeV}$ will
be adopted in the following. Notice please, that the first term in 
(\ref{averendif}) stands for the kinetic energy $E^{\rm traj}_{\rm
kin}= M\varpi^2q_{sc}^2 \zeta^2 /2$ one would obtain in 
a mere trajectory calculation. This is easily verified from the solution
\bel{tranewt}
q_{sc}=q(\tau) \approx {P_0\over 2M\varpi\sqrt{1+\eta^2}}
\exp(\varpi(\sqrt{1+\eta^2} - \eta)\tau)
\ee
of Newton's equation for $q(t=0)=0$, only assuming the
saddle to scission time $\tau$ to be sufficently large to have
$\varpi\sqrt{1+\eta^2} \tau \gg 1$. In (\ref{averendif}) all the other
terms represent the effects of the fluctuating force. 

Before turning to discuss numerical evaluations of the expressions just
presented it may be worth while to comment on their physical relevance.
Firstly, we like to stress that by using the stationary solution 
$d_I(q,P)$ the results become {\it insensitive} to initial conditions, 
like those on top of the barrier one would need to invoke in a time
dependent picture. Please notice that the only uncertainty left in 
$d_I(q,P)$ is the multiplicative factor hidden in the current $j$ which
drops out when calculating the $E_{\rm kin}$ and $\sigma^2_{\rm kin}$
according to (\ref{averen}) and (\ref{fluc}), respectively. For the
situation to which this stationary solution $d_I(q,P)$ of the inverted
oscillator commonly is applied to (c.f.\cite{kram}), this $j$ can be
said to stand for the decay rate out of the potential miminum.
Eventually, one would then like to have this minimum to be well
pronounced, in the sense of having the barrier height be large compared
to $T$, more precisely to that temperature the system has on its way
towards the saddle. However, as demonstrated in \cite{scheuhof} and
\cite{nixshsv} the $d_I(q,P)$ may as well be understood to result from
integrating a $t$-dependent distribution $d(q,P,t)$ over time $t$. It
is not difficult to understand that such a time integrated function
solves an equation like (\ref{traneq}) with the left hand side put
equal to zero.  Moreover, as shown in \cite{nixshsv}, already for quite
small damping rates $\eta$ (called $\gamma$ in \cite{nixshsv}) this
time-integrated distribution shows a relaxational behavior to Kramers
solution, if considered in its dependence on $q$. In this spirit,
formulas (\ref{averen}) to (\ref{enerfluctdif}) may be understood in
the following way. Provided that the scission point $q_{sc}$ does not
lie too closely to the position of the barrier top, we may assume the
$d_I(q,P)$ to adequately portray the momentum distribution of an actual
fission process.  Formulas (\ref{averendif}) and (\ref{enerfluctdif})
then measure the kinetic energy and its fluctuation for the
distribution one obtains after summing up all events leading to
fission.

In Fig.1~we show the kinetic energy as function of $\eta$ for
different temperatures as calculated from  (\ref{averendif}). 
\begin{figure}[htb]
\centerline{
\rotate[r]
{\epsfysize=13cm
\epsffile[95 78 549 669]{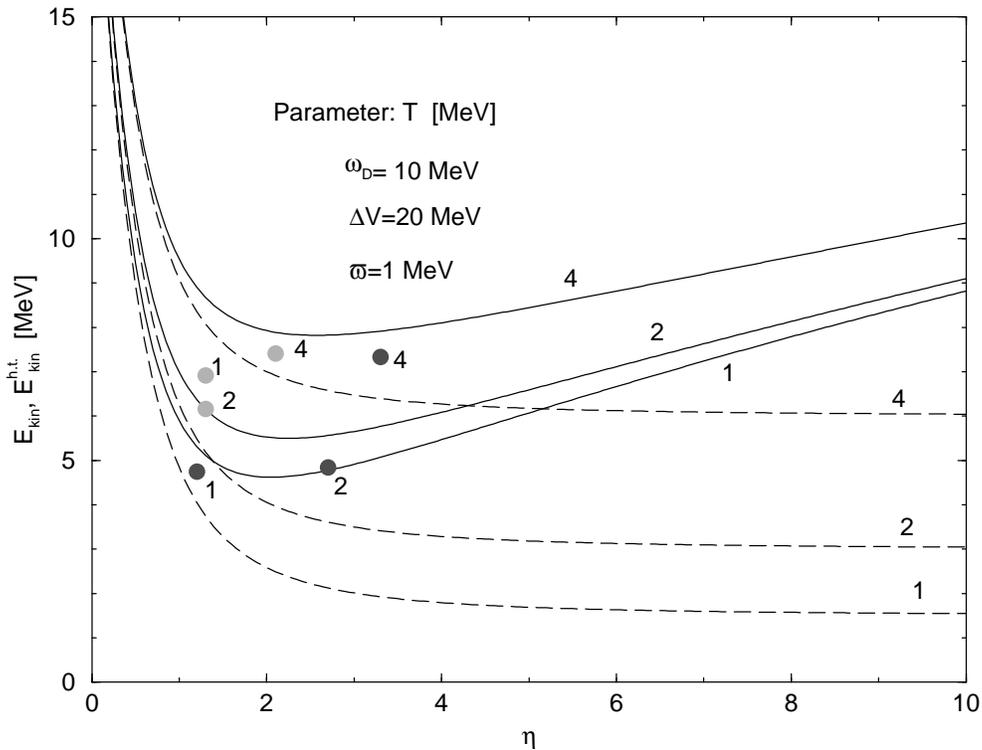}}}
\caption{
Average kinetic energy as function of $\eta$ for different 
temperatures $T$. Fully drawn curves include quantum
effects, dashed ones refer to Kramers' solution. 
Shaded dots give results based on the values for
$\varpi$ and $\eta$ (at given T) as obtained in 
microscopic computations of [7]%\cite{yaivho} 
(dark shaded) and [8]%\cite{ivhopaya} 
(light shaded). (In this and the
following figures we use units of $\hbar=1$). 
}
\end{figure}
Fully drawn lines correspond to the quantal case and dashed ones
to the high temperature limit, thus corresponding to Kramers' equation.
As the figure demonstrates, quantum effects increase with damping and
they amplify the average kinetic energy. As a matter of fact, the
latter is seen to attain quite large values in any case, if compared to
the value of $\Delta V$. Obviously, this is an effect of the
fluctuating force, as friction alone acts to diminish the velocity and
thus the kinetic energy. This feature is demonstrated explicitly in
Fig.2. There the ratio $E^{\rm traj}_{\rm kin} /E_{\rm kin}$ is shown
(with the $E^{\rm traj}_{\rm kin}$ introduced below
(\ref{enerfluctdif})). 
\begin{figure}[htb]
\centerline{
\rotate[r]
{\epsfysize=13cm
\epsffile[95 61 549 669]{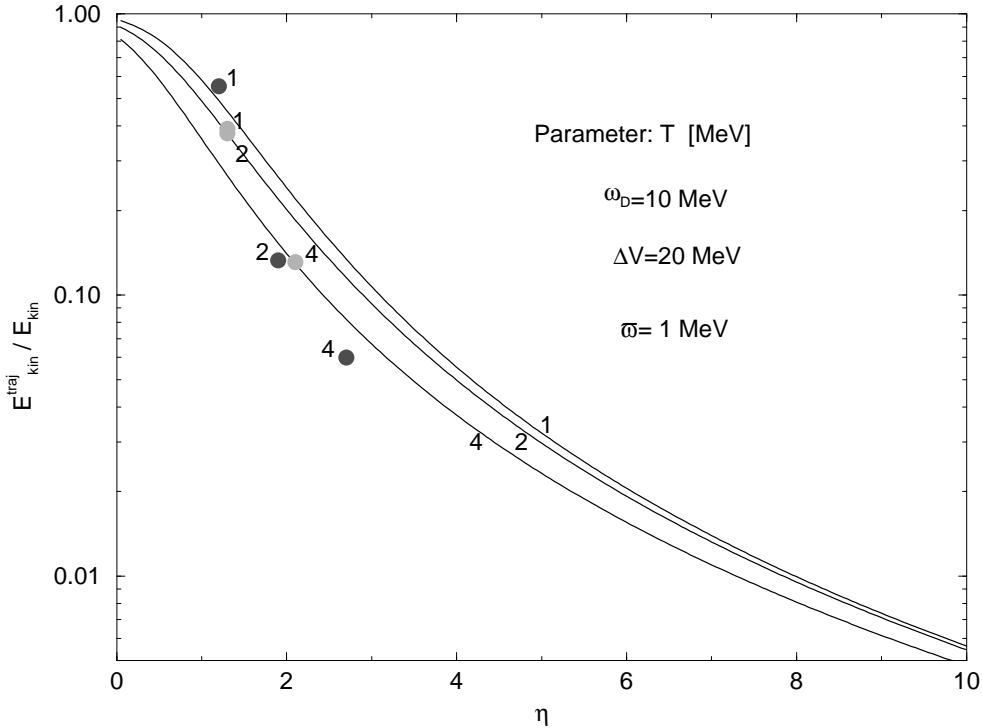}}}
\caption{
Ratio of the kinetic energies, $E^{\rm traj}_{\rm kin}$ and 
$E_{\rm kin}$, calculated, respectively, without and with
fluctuating force. The dots are specified as in Fig.1.
}
\end{figure}
In both figures shaded circles indicate results
obtained by using the transport coefficients for average motion as
found in microscopic computations of \cite{yaivho} (dark shaded) and
\cite{ivhopaya} (light shaded).  Fig.2~demonstrates clearly that
trajectory calculations may grossly underestimate the size of $E_{\rm
kin}$.

The very fact of the big influence of the fluctuating force hints at
the importance of statistical fluctuations of the kinetic energy
itself, which may be calculated according to (\ref{enerfluctdif}).  In
Fig.3~we show the square root of the variance $\sigma$
divided by the kinetic energy  $E_{\rm kin}$, 
both calculated at the scission point. 
\begin{figure}[htb]
\centerline{
\rotate[r]
{\epsfysize=13cm
\epsffile[104 75 549 669]{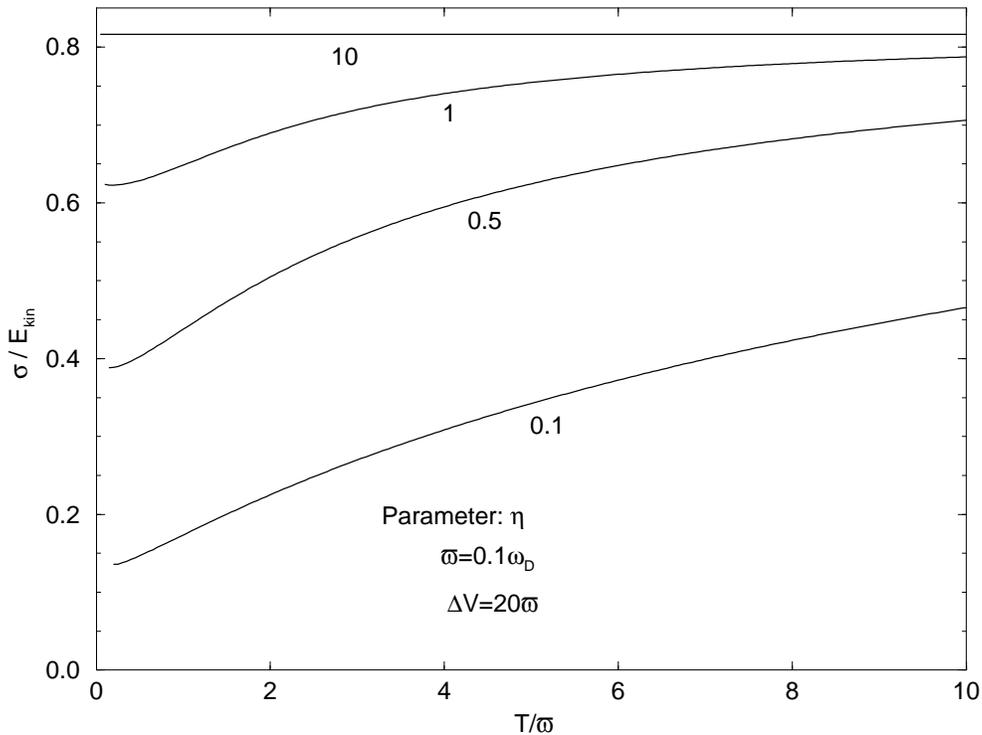}}}
\caption{
Variance $\sigma$ over average kinetic energy $E_{\rm kin}$ 
as function of $T/\varpi$, for different values of $\eta$.
}
\end{figure}
For large damping and large
temperatures this ratio comes close to the limiting value $\sqrt{2/3}$.
When the effective damping rate $\eta$ is somewhat larger than $1$ this
value of $\sqrt{2/3}$ is reached for practically all $T$. Incidentally,
we may note that this ratio $\sigma / E_{\rm kin}$ can be expected less
sensitive to the Drude frequency $\omega_D$ than the individual
quantities, simply because both of the quantities, $\sigma$ and 
$ E_{\rm kin}$, change with $\omega_D$ alike.

Next we like to work out more explicitly the size of quantum effects. 
In Fig.4~we plot for both the kinetic energy and its fluctuations, 
the ratio of the values in the quantal case to the corresponding 
ones in the high-T limit (Kramers' case).  
\begin{figure}[htb]
\centerline{
\rotate[r]
{\epsfysize=13cm
\epsffile[95 89 549 667]{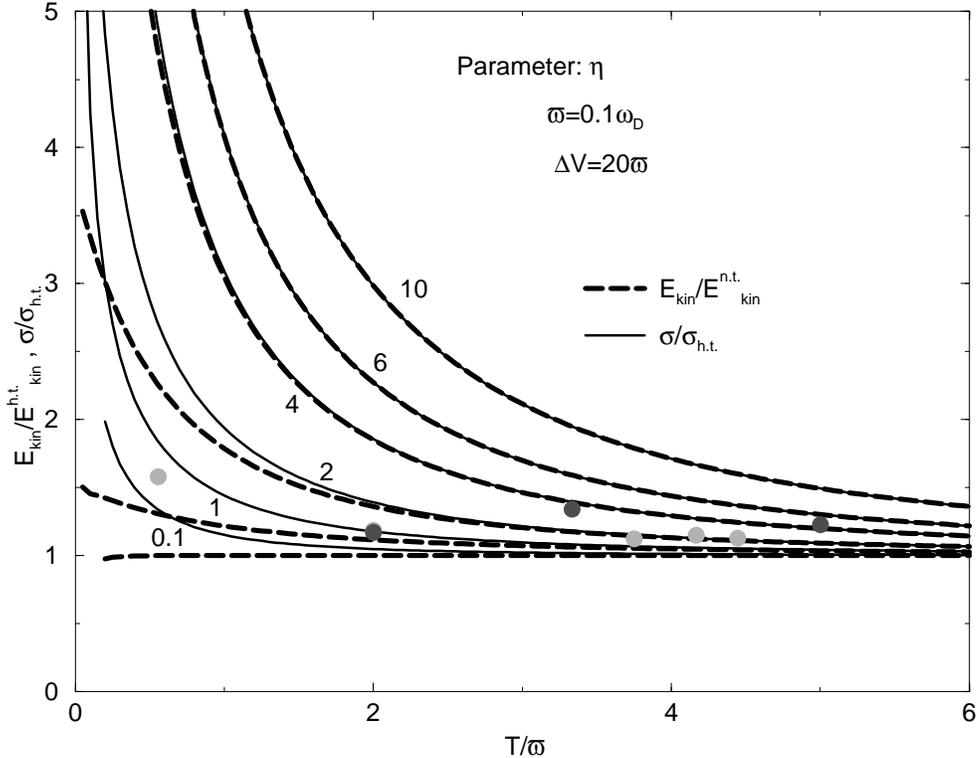}}}
\caption{
Ratios of kinetic energy and its variance to their values in the 
high temperature limit, both as functions of $T/\varpi$,
again for different $\eta$.
}
\end{figure}
As seen from
the figure, these ratios may take on quite large values and they increase
with increasing damping. Thus Fig.4~agrees with the
observation made in Fig.1~and confirms our conjecture raised
earlier: Whenever the collective momentum is involved quantum effects
get larger with increasing damping. Like before we have again indicated
by dots the range one would expect for these values on the basis of
the microscopic computations of \cite{yaivho} and \cite{ivhopaya}. 
Notice, please, that in these computations the coefficients $\eta$
and $\varpi$ have been evaluated as function of $T$. For the curves
shown in all the figures these coefficients have been varied as free
parameters. This will facilitate comparison with other
theoretical models and with results deduced from experiments.
Indeed, it may be said that there is experimental evidence (see e.g.
\cite{hirorev} and \cite{pauthoe}) for much larger damping coefficients
than found in \cite{yaivho} and \cite{ivhopaya}. The authors of 
\cite{pauthoe} need values of as much as $\eta=10$ to cope with
findings in experiments where fission of heavy nuclei is observed 
accompanied by GDR $\gamma$ rays, at temperatures not larger than about
$2 {\rm MeV}$. Applying macroscopic pictures to evaluate the transport
coefficients, like the wall formula for friction, irrotational flow for
inertia and the liquid drop model for the stiffness one would find
similar values \cite{pompc} (for $^{224}{\rm Th}$ at $T=2 \;{\rm MeV}$
$\eta$ becomes $\approx 7.5$). If one looks at lighter systems where
large angular momenta are needed to find fission events with some
finite chance \cite{pobaridi}, the barrier becomes quite small and
broad, thus leading to smaller values of $\varpi$ and hence to larger
values of $\eta$ \cite{pompc}. As seen from our figures, this hints not
only at the importance of the statistical fluctuations as such, but
also at the necessity of calculating them with the quantal diffusion
coefficients. Most likely this may again modify the interpretation of
experimental results. In any case, it is probably fair to say that
still some work is to be done before more conclusive statements can be
made about the size of the transport coefficients. In this context one
may mention that there are indications from other experiments that
$\eta$ might be smaller, indeed, than the macroscopic picture requires
\cite{jolestpc}. 

So far in this paper we have been looking at cases which explicitly
involve the momentum distribution in one way or other. For the sake of
completeness we should like to take up once more the question of the
influence of statistical fluctuations on the time $\tau$ it takes for
the system to move from saddle to scission. In \cite{hofnix} the
following formula had been derived, based on Kramers' equation:
\bel{timess}
\tau= {\tau_0 \over \sqrt{1+\eta^2}-\eta}\equiv
{2{\cal R}(\sqrt{M\varpi^2q_{sc}^2/2T})
\over \varpi (\sqrt{1+\eta^2}-\eta)}
\ee 
with ${\cal R}(x) $ being the Rosser function
\bel{rosser}
{\cal R}(x) =\int_0^x \exp(y^2)dy\,\int_y^\infty \exp(-z^2)dz
\ee 
In the quantum case the argument of the ${\cal R}$ in (\ref{timess})
would get the additional factor $\sqrt{T/C\Sigma^{\rm eq}_{qq}}$
\cite{hofing}, but as can be seen from Fig.2 of this reference this
modification may safely be neglected for $T/\hbar \varpi > 0.5$. For
that regime Fig.5 shows the $\varpi \tau_0$ as function of $\Delta V/T =
M\varpi^2q_{sc}^2/2T$. 
\begin{figure}[htb]
\centerline{
\rotate[r]
{\epsfysize=13cm
\epsffile[104 73 549 669]{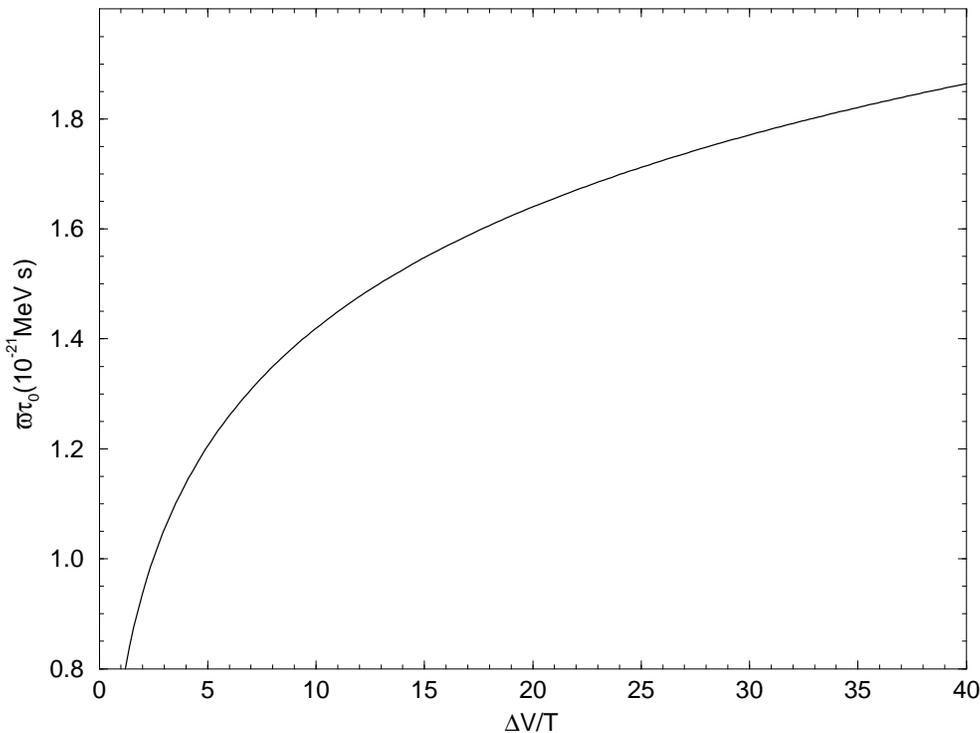}}}
\caption{
The mean saddle-to-scission time $\tau_0$ (scaled by $\varpi$) as
function of the difference $\Delta V$ of the potential energy (measured in
units of $T$). 
}
\end{figure}
As the $\hbar \varpi$ is of the order of $1$
MeV, the $\tau_0$ is seen to lie in the range of $1~-~2\times 10^{-21}
{\rm s}$. It is interesting to compare the result (\ref{timess}) with
the one one would get if the system would move along a trajectory
starting on top of the barrier with the momentum $P_0$ (i.e. including
disipation but discarding the fluctuating force). As seen from
(\ref{tranewt}) one gets a formula like (\ref{timess}) but with the 
$2{\cal R}(\sqrt{M\varpi^2q_{sc}^2/2T})$ replaced by  
$\ln(2M\varpi q_{sc}\sqrt{1+\eta^2}/P_0)$. To make
the analogy to (\ref{timess}) even closer one may estimate the initial
momentum by associating $P_0^2/2M$ to an average, thermal kinetic
energy on top of the barrier. One might be tempted to use for the
latter $E_{\rm kin}=T/2$, the value given by the equipartition theorem.
A numerical evaluation shows that in this case 
one would overestimate the $\tau_0$ by about 10\% (for $\eta=1$) to 
50 \% (for $\eta=10$).  However, as we may learn from
(\ref{averendif}), for the inverted oscillator the stationary solution
suggests a larger value of the average kinetic energy. Putting there
the $q_{sc}=0$ one obtains (in the high-T limit) $E_{\rm kin}=T(1+\eta
\zeta)$, which for zero damping gives twice the value of the
equipartition theorem.  This modification would improve the results
slightly.

Finally, we wish to add some remarks on the variation with $q$. It is
evident from the discussion given above that under certain
circumstances the Kramers-Ingold solution may be applied for regions
before scission.  It may thus be used for performing averages over
quantities which simply depend on the collective variable. Prime
examples are the evaporation probabilities for light particles
\cite{jolest} and $\gamma$-rays (see \cite{pauthoe}). Adopting the same
normalization as in (\ref{averen}, \ref{fluc}) one gets as weighting
factor the current $j$ which according to (\ref{flux}) is constant.
This implies that for the motion from saddle to scission an average in
the coordinate $q$ simply reduces to the algebraic one.

{\it Acknowledgements:} The authors like to thank K. Pomorski for
valuable suggestions, and they gratefully acknowledge financial support
by the Deutsche Forschungsgemeinschaft.

\end{document}